\documentclass[aps,floatfix,twocolumn,floats,superscriptaddress]{revtex4-1}

\usepackage[T1]{fontenc}
\usepackage{times}
\usepackage{graphicx}
\usepackage{amsmath}
\usepackage{amssymb}
\usepackage{xcolor}
\usepackage{url}
\usepackage{braket}
\usepackage{textcomp}

\usepackage{ulem}
\newcommand{\ham}{\hat{\mathcal{H}}}

\newcommand{\quotes}[1]{``#1''}

\begin{document}
\title{
Tailoring CIPSI expansions for QMC calculations of electronic excitations: the case study of thiophene
}
\author{Monika Dash}
\affiliation{MESA+ Institute for Nanotechnology, University of Twente, P.O. Box 217, 7500 AE Enschede, The Netherlands}
\author{Saverio Moroni}
\email{moroni@democritos.it}
\affiliation{CNR-IOM DEMOCRITOS, Istituto Officina dei Materiali, and SISSA Scuola Internazionale Superiore di Studi Avanzati, Via Bonomea 265, I-34136 Trieste, Italy}
\author{Claudia Filippi}
\email{c.filippi@utwente.nl}
\affiliation{MESA+ Institute for Nanotechnology, University of Twente, P.O. Box 217, 7500 AE Enschede, The Netherlands}
\author{Anthony Scemama}
\email{scemama@irsamc.ups-tlse.fr}
\affiliation{Laboratoire de Chimie et Physique Quantiques, Universit\'e de Toulouse, CNRS, UPS, France}

\begin{abstract}

The perturbatively selected configuration interaction scheme (CIPSI) is
particularly effective in constructing determinantal expansions for
quantum Monte Carlo (QMC) simulations with Jastrow-Slater wave functions: fast and smooth
convergence of ground-state properties, as well as balanced descriptions
of ground- and excited-states of different symmetries have been reported.
In particular, accurate excitation energies have been obtained 
by the pivotal requirement of using CIPSI expansions with similar 
second-order perturbation corrections for each state, 
that is, similar estimated errors with respect to the full configuration interaction limit.
Here we elaborate on the CIPSI selection criterion for excited states of the same symmetry 
as the ground state, generating expansions from a common orbital set.
Using these expansions in QMC as determinantal components of Jastrow-Slater wave functions, we compute 
the lowest, bright excited state of thiophene, which is challenging due to its 
significant multireference character.  The resulting vertical excitation energies
are within 0.05~eV of the best theoretical estimates, already with expansions of only a few thousand
determinants.  Furthermore, we relax the ground- and excited-state structures following the corresponding
root in variational Monte Carlo and obtain bond lengths which are accurate to better than 0.01~\AA.  
Therefore, while the full treatment at the CIPSI level of this system would be quite demanding, 
in QMC we can compute high-quality excitation energies and excited-state structural 
parameters building on affordable CIPSI expansions with relatively few, well chosen determinants.


\end{abstract}

\maketitle

\section{Introduction}

The accurate description of photoinduced phenomena relies on the balanced treatment
of the multiple electronic states involved in the excitation process.  Recently,
we have demonstrated the ability of quantum Monte Carlo (QMC) methods to yield
chemically accurate ground- and excited-state structures as well as vertical and
adiabatic excitation energies for small, prototypical molecules~\cite{dash2018,dash2019}.
We used wave functions of the Jastrow-Slater form where the determinantal component
was generated in an automatic manner with the configuration interaction using a
perturbative selection made iteratively (CIPSI) approach.  If the expansion was
then fully optimized together with the Jastrow factor in
QMC~\cite{filippi2016,assaraf2017}, a handful of CIPSI determinants was found to
be sufficient to provide well converged geometries and excitation energies.

Importantly, we showed~\cite{dash2019,cuzzocrea2020} that a balanced QMC description
of the ground and excited states also at different geometries could be achieved
by generating CIPSI expansions characterized by the same second-order perturbation
(PT2) energy correction, that is, the same ``error'' with respect to the full CI
limit.  
Furthermore, these ``iso-PT2'' expansions were found to have similar values of the CI variance
which is another useful measure of the error of a CIPSI wave function.

If the excited states investigated belong to a different symmetry class than the
ground state, one can perform the expansions either separately, stopping when the
same target PT2 energy correction or CI variance is reached~\cite{cuzzocrea2020}, or concurrently
with a common set of orbitals. In the latter case, the difference in symmetry and
the relatively dominant single-reference character of the states investigated aided
the CIPSI selection, and we heuristically found that a rather straightforward
common selection criterion closely approaches the iso-PT2 condition~\cite{dash2019}.

If the states are of the same symmetry, separate expansions do not guarantee
orthogonality, and the preferred route is the use of concurrent CIPSI expansions
on a common set of orbitals.  In this case, however, rendering their balanced
description is more difficult and the simple selection scheme employed in
Ref.~\cite{dash2019} proves inadequate, especially if some of the relevant states
are strongly multi-configurational.

Here, we propose a simple and effective modification of the selection criterion
to enable the construction of nearly iso-PT2 CIPSI expansions for multiple states
of the same symmetry, and illustrate the scheme on the challenging case of thiophene.
This molecule forms the backbone of a class of $\pi$-conjugated donor polymers in
organic solar-cell
devices~\cite{roncali1992,zhang2011,bian2012,xu2013,meerheim2014,duan2017} and the
computation of its electronic excited states has been the subject of several
theoretical
investigations~\cite{wan2001,kleinschmidt2002,pastore2007,salzmann2008,stenrup2012,holland2014,prlj2015,prlj2015b,kolle2016,schnappinger2017}.
In particular, the accurate prediction of the lowest, bright excited state of
thiophene is difficult for traditional {\it ab-initio} methods, with different
levels of theory spanning a range of about 0.5~eV as illustrated below.  The
multireference character of this state calls in fact for the use of highly-correlated
electronic structure methods and its inherent complexity renders this a perfect
test case for our modified CIPSI selection approach.

By simply assigning a higher weight in the CIPSI selection criterion to the state
showing a slower convergence, we succeed in generating CIPSI
expansions for the ground and excited states of thiophene fulfilling the basic
condition of similar PT2 energy corrections and CI variances.  
The resulting Jastrow-Slater wave functions yield variational Monte Carlo (VMC)
vertical excitation energies which are lower and in closer agreement with the
diffusion Monte Carlo (DMC) counterparts than those provided by the default selection
scheme.  Furthermore, we obtain converged estimates of the QMC vertical excitation
energy already with compact expansions containing a few thousand determinants, and
our best estimates are within 0.05~eV of the reference coupled cluster (CC) value.
Finally, we compute the optimal ground- and excited-state geometries
in VMC, following the relevant root and generating CIPSI expansions with
similar PT2 corrections for both states along the optimization path.
The optimal VMC structural parameters are in excellent agreement with the CASPT2
or CC estimates, namely, within 0.01 \AA\ for the bond lengths and, in the excited
state, 1$^\circ$ for the bond angles.

This article is organized as follows: We describe the modified CIPSI selection
procedure for multiple states of the same symmetry in
Section~\ref{sec:chap5-methods}, and present the computational details in
Section~\ref{sec:compdetails}. The construction procedure of the wave functions
with the modified selection scheme is detailed in
Section~\ref{sec:wf-construct}. We present the QMC vertical excitation energies
in Section~\ref{sec:chap5-vexc} and the optimal ground- and excited-state VMC
structures in Section~\ref{sec:chap5-optgeo}. We conclude in
Section~\ref{sec:chap5-summary}.

\section{Methods}
\label{sec:chap5-methods}

In this work, we consider excited states that are not the lowest in their
symmetry class. In the QMC calculations, we describe them together with the
lower-energy states of the same symmetry via a set of Jastrow-Slater wave
functions with different CI coefficients but the same Jastrow and orbital
parameters:
\begin{equation}
\Psi_n = {\cal J}\sum_{k=1}^{N_{\rm det}}c_k^n D_k\,
\label{eq:j-slater-sa}
\end{equation}
where $N_{\rm det}$ is the total number of determinants and the index $n$
denotes an electronic state. We use a Jastrow factor which describes
electron-nucleus and electron-electron correlations (${\cal J}_{\rm 2-body}$),
and guarantees that Kato's cusp conditions are satisfied at the inter-particle
coalescence points.

To ensure that the common non-linear (Jastrow and orbital) parameters offer a
reasonable description of all states of interest, we optimize them by
minimizing the state-average (SA) energy~\cite{filippi2009} defined as
\begin{equation}
E^{\rm SA} = \sum_{n=1}^{N_{\rm states}} w_n^{\rm QMC}\frac{\langle\Psi_n|\ham|\Psi_n\rangle}{\langle\Psi_n|\Psi_n\rangle}\,,
\label{eq:en-sa}
\end{equation}
where the weights, $w_n^{\rm QMC}$, sum up to one and are kept fixed during the
optimization.  Orthogonality between the states is maintained through the
(linear) CI coefficients whose optimal values are obtained by solving a
generalized eigenvalue problem in the basis of the determinants multiplied by
the Jastrow factor.

As described in Ref.~\cite{cuzzocrea2020}, we alternate a number of
optimization steps of the non-linear parameters with a step of optimization of
the linear coefficients. For the former, we follow the down-hill gradient of
the SA energy in a scheme inspired by the stochastic reconfiguration approach
for a single state~\cite{sorella2007}, and solve the relevant equations in a
low-memory conjugate-gradient implementation~\cite{neuscamman2012}.  For the
latter, we use a memory-efficient Davidson diagonalization method that allows
the computation of the lowest energy eigenvalues without explicit construction
of the entire Hamiltonian and overlap
matrices~\cite{neuscamman2012,sabzevari2020}.  Combining this optimization
scheme with the fast generation of the quantities needed in the QMC
estimators~\cite{filippi2016,assaraf2017}, we can optimize QMC wave functions
for ground and excited states containing large determinantal expansions and
several thousand parameters.

To construct the determinantal component of these wave functions, we employ an
improved CIPSI approach that allows us to iteratively select the most important
determinants required for the balanced description of multiple electronic
states. Starting from an initial reference subspace, $\cal S$, given by the
union of determinants describing the states of interest, an external
determinant $\ket{\alpha}$ is selected based on its weighted second-order
perturbation (PT2) energy contribution obtained via the Epstein-Nesbet
partitioning of the Hamiltonian~\cite{epstein1926,nesbet1955},
\begin{equation}
e_{\alpha} = \sum_{n=1}^{N_{\rm states}} w_n\delta E_{\alpha,n}^{(2)},
\label{eq:e-pt2}
\end{equation}
where
\begin{eqnarray}
\delta E_{\alpha,n}^{(2)} = \frac{|\langle\alpha|\ham|\Psi_n^{\rm CIPSI}\rangle|^2}{\langle\Psi_n^{\rm CIPSI}|\ham|\Psi_n^{\rm CIPSI}\rangle - \langle\alpha|\ham|\alpha\rangle}\,,
\label{eq:en-pt2}
\end{eqnarray}
and $\Psi_{n}^{\rm CIPSI}$ is the current normalized CIPSI wave function for state
$n$. In this partitioning scheme, the first-order energy correction is zero by
definition. The determinant $\ket{\alpha}$ is added to $\cal S$ if its energy
contribution $e_\alpha$ is higher than a threshold, and the threshold is automatically
adjusted so that the number of determinants in $\cal S$ is increased by a certain
percentage at every iteration.

We have recently shown that the use of iso-PT2 CIPSI expansions results in a
balanced description of the relevant states when complemented by a Jastrow factor
and fully optimized in QMC, these expansions were found to yield accurate QMC
estimates of the excitation energies for relatively small numbers of
determinants~\cite{dash2019}.

When interested in the lowest-energy states of different symmetries, one can in
principle perform the expansion separately for each state until the corresponding
perturbation energy contribution is equal to a target value.  Alternatively, one
can generate the expansions using one set of orbitals for all states and enlarging
the union space, $\cal S$, through a single threshold. Following this last scheme,
we were able to obtain matched PT2 corrections~\cite{dash2019} by simply choosing
the weights in the selection step (Eq.~\ref{eq:e-pt2}) as:
\begin{eqnarray}
w_n = w_n^{\rm cmax} = \frac{1}{{\rm max}(c_{k,n}^2)}\,,
\label{eq:w_simple}
\end{eqnarray}
where the index $k$ runs over all determinants in the current $\Psi_{n}^{\rm
CIPSI}$. Such a choice follows Ref.~\cite{angeli1997}, with modifications due to
the fact that we perform the CIPSI selection in the basis of determinants and not
of configuration state functions (CSFs).  The resulting expansions for states of different symmetry
have different sizes since they do not share any common determinant (determinants
of a given symmetry have zero coefficients in the expansion for a state belonging
to a different symmetry class).

Here, we are interested in states of the same symmetry expanded on the same set
of determinants, where a CSF may have non-zero and non-negligible
overlap with all states of interest. In this case, we find that the use of the
simple weights (Eq.~\ref{eq:w_simple}) does not yield iso-PT2 expansions. To improve
this balance, we explore a simple modification to the selection scheme where we
multiply the weights of Eq.~(\ref{eq:w_simple}) by user-given ``state-average''
weights, $w_n^{\rm SA}$, i.e.
\begin{equation}
w_n = w_n^{\rm cmax}\times w_n^{\rm SA}\,.
\label{eq:wSA}
\end{equation}

In addition to the PT2 correction, we investigate the behavior of the CI variance,
$\sigma^2_{\text{CI}}$, which is defined as the variance of the full CI (FCI)
Hamiltonian:
\begin{align}
\sigma^2_{\text{CI}} (\Psi_n^{\rm CIPSI}) =
\sum_{\alpha \in \text{FCI}} &
\langle \Psi_n^{\rm CIPSI} |\ham| \alpha \rangle
\langle\alpha|\ham|\Psi_n^{\rm CIPSI}\rangle   \nonumber \\
& - \langle \Psi_n^{\rm CIPSI} |\ham| \Psi_n^{\rm CIPSI} \rangle ^ 2\,,
\end{align}
and goes to zero as the CIPSI wave function approaches the FCI limit. Since the
CI variance is also an indicator of the quality of the CIPSI wave function, one
can match the CI variances of the states of interest together with the PT2 energy
correction or as an alternative to the PT2 criterion.  We note that the CI variance
should not be confused with the QMC variance which is defined in terms of the exact
Hamiltonian.

\section{Computational details}
\label{sec:compdetails}

All QMC calculations are performed with the program package CHAMP~\cite{Champ}.
We employ the Burkatzki-Filippi-Dolg (BFD) scalar-relativistic energy-consistent
Hartree-Fock (HF) pseudopotentials and correlation consistent Gaussian basis sets
that have been specifically constructed for these
pseudopotentials~\cite{burkatzki2007,BFD_H2013}.  When unclear, we append the
\quotes{(BFD)} suffix to these basis sets to avoid
  confusion with the corresponding all-electron basis sets.
For most test QMC calculations, we use a minimally augmented double-$\zeta$
(maug-cc-pVDZ) basis set, where the basis on the heavy atoms are augmented
with $s$ and $p$ diffuse functions.  Final calculations are computed with the fully
augmented double (aug-cc-pVDZ) and triple (aug-cc-pVTZ) basis
sets.  All diffuse functions are obtained from the corresponding all-electron
Dunning's correlation-consistent basis sets~\cite{kendall1992}.
We employ a two-body Jastrow factor including electron-electron and electron-nucleus
correlation terms~\cite{Jastrow}.

We optimize all parameters (Jastrow, orbital, and CI coefficients) of the Jastrow-Slater wave function in VMC as described
above. We employ equal weights $w_n^{\rm QMC}$ in the state-average energy of
Eq.~(\ref{eq:en-sa}) and a guiding wave function $\Psi_{g}^{2} = \sum_{n}|\Psi_{n}|^{2}$
in the sampling to ensure a reasonable overlap with all states of
interest~\cite{filippi2009}.  In the VMC geometry optimization, we relax the
structure without symmetry constraints and simply follow the path of steepest
descent for the root of interest by appropriately rescaling the interatomic forces
and using an approximate constant diagonal Hessian. After convergence, we perform
40 additional optimization steps to estimate the optimal average structural
parameters.  In the DMC calculations, we treat the pseudopotentials beyond the
locality approximation using the $\rm T$-move algorithm~\cite{casula2006a} and
employ an imaginary time step of 0.02 a.u.\ with single-electron moves. This time
step yields DMC excitation energies converged to better than 0.01~eV also for the
simplest wave function employed here (see S2).

We carry out the CIPSI calculations with Quantum Package~\cite{garniron2019} using
orbitals obtained from complete active space self-consistent field (CASSCF)
calculations performed with the program GAMESS(US)~\cite{schmidt1993,gordon2005},
correlating six electrons in five $\pi$ orbitals in a minimal CAS(6,5).  The CIPSI
expansions are constructed to be eigenstates of $\hat{S}^2$ and are mapped into
the basis of CSFs, effectively reducing the number of optimization parameters in
VMC. The PSI4 package~\cite{parrish2017} is used to compute the reference ground
state geometry and the Dalton package~\cite{dalton2014,Daltonversion} is used to
compute the vertical excitation energies within the iterative approximate coupled
cluster singles, doubles, and triples (CC3) approach. We perform
the {\it n}-electron valence perturbation theory (NEVPT2) calculations using the 
Molpro 2019.2 code~\cite{werner_2012,MOLPRO}.

All vertical excitation energies are computed on the fixed ground-state
structure optimized at the CC3 level with an all-electron aug-cc-pVTZ
basis and the frozen-core (FC) approximation. 
Unless explicitly stated, the calculations presented below are
computed with the BFD pseudopotentials and the corresponding basis sets,
irrespective of the level of theory.

\section{Results and Discussion}

We focus here on the lowest-lying, bright $\pi \rightarrow \pi^{*}$ singlet excited
state of the thiophene molecule (C$_{4}$H$_{4}$S).  The ground-state structure of
thiophene has C$_{2v}$ symmetry and the ground and the targeted excited state
belong to the A$_1$ irreducible representation.  The accurate computation of this
excited state is challenging because of its multi-reference character: in a CASSCF
calculation with the minimal active space correlating six electrons in five $\pi$
orbitals, one finds that the two dominant transitions, HOMO$-$1 $\rightarrow$ LUMO
and HOMO $\rightarrow$ LUMO$+$1, account for almost 60\% and 20\% of the wave
function, respectively, while higher-order double excitations with four unpaired
electrons make up for another 7\%. The ground state, on the other hand, is single
reference, dominated by the HF determinant which alone accounts for more than 90\%
of the wave function in the same calculation.

\begin{figure}[t]
\centering
\includegraphics[width=1.0\columnwidth]{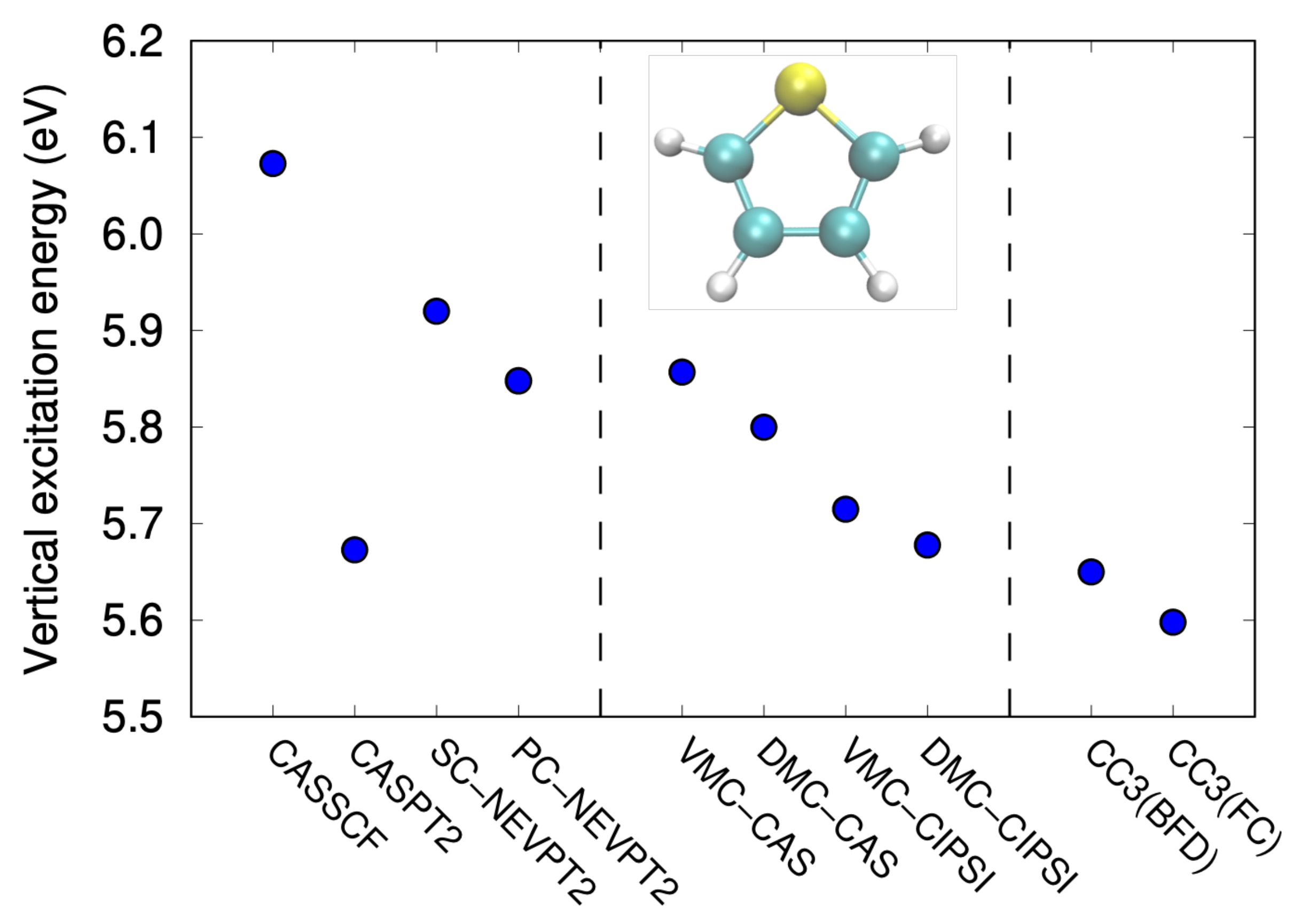}
\caption{
Vertical excitation energy of thiophene for the $\pi\to\pi^*$ singlet transition
to the lowest, bright excited state, computed with different approaches. All
calculations are performed with the BFD pseudopotentials and the corresponding
aug-cc-pVTZ basis set with the exception of the all-electron CC3/aug-cc-pVTZ (FC)
calculation.  We also include schematic representation of thiophene, where yellow,
blue, and white denote sulfur, carbon, and hydrogen, respectively. 
}
\label{fig:thiophene-methods}
\end{figure}

The nature of this excited state leads to difficulties in estimating the corresponding
vertical excitation energy, which we find to span a range of about 0.5~eV, between
5.60 and 6.07~eV, across different levels of theory as illustrated in
Fig.~\ref{fig:thiophene-methods}.  Time-dependent density functional theory (TDDFT)
in combination with different exchange-correlation functionals incorrectly places
a $\pi \rightarrow \pi^{*}$ state of B$_{2}$ symmetry lower than the A$_1$ state
by about 0.1~eV or less~\cite{prlj2015b}.  All tested wave function methods (CASPT2,
NEVPT2, QMC, and CC3), instead, identify our state of interest as the
energetically lowest singlet excited state, in qualitative agreement with
experimental observations \cite{rolf1977,norden1978,igarashi1981,holland2014};
however, they yield very different excitation energies.

Estimating the FCI excitation energy of thiophene at the CIPSI level is challenging
because of the size of the FCI space. Consequently, CIPSI calculations are limited
to relatively small basis sets and yield extrapolated energy differences characterized
by uncertainties as large as 0.08~eV~\cite{Veril2021}.  Since, within a small basis
set, CC3 was found to yield an excitation energy of thiophene compatible with the
FCI estimated by extrapolating very large CIPSI calculations~\cite{Veril2021}, one
can use this level of theory to compute the reference excitation energy for this
state of thiophene.  Furthermore, the CC3/aug-cc-pVTZ value was shown to differ
from the complete basis set limit at the same level of theory by only
0.02~eV~\cite{loos_2020}.  Therefore, since we use a slightly different geometry
than the one of Refs.~\cite{Veril2021} and~\cite{loos_2020} and the QMC calculations
are done with pseudopotentials, we compare our QMC results to the CC3/aug-cc-pVTZ(BFD)
excitation energy computed with our geometry, which we believe to be an accurate
estimate of the exact vertical excitation energy for our study.


\subsection{Modified selection criterion for CIPSI}
\label{sec:wf-construct}

\begin{figure}[htbt!]
\includegraphics[width=1.0\columnwidth]{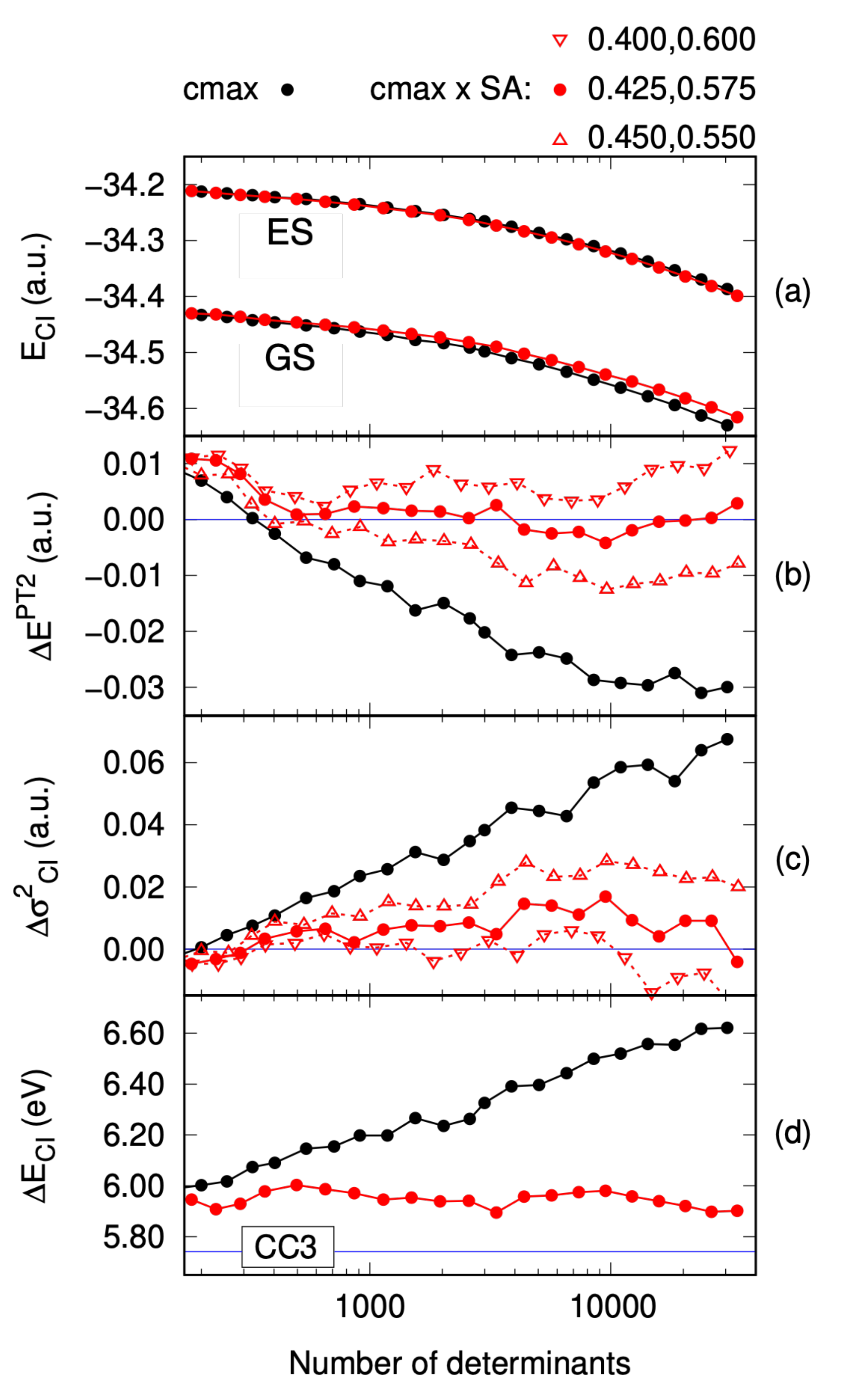}
\caption{
CI results for the ground- (GS) and excited-state (ES) of thiophene with CIPSI
expansions of increasing size: total energy (a); difference of the ground- and
excited-state PT2 energy corrections $\Delta E^{PT2}$ (b) and CI variances
$\Delta\sigma^2_{CI}$ (c); vertical excitation energy $\Delta E_{CI}$ (d).  We
employ two reweighting schemes in the selection criterion, namely, $w_n^{\rm cmax}$
and $w_n^{\rm cmax}\times w_n^{\rm SA}$, and different choices of state-average
weights, $w_n^{\rm SA}$.  The maug-cc-pVDZ basis set is used.
}
\label{fig:method-compare}
\end{figure}

For the ground and excited states of thiophene, we construct CIPSI expansions
of increasing length from subsequent CIPSI iterations. To test the different selection
criteria, we use the maug-cc-pVDZ basis set and the common set of CAS(6,5) orbitals.

We first adopt the simple reweighting scheme (Eq.~\ref{eq:w_simple}) that we
successfully used for states of different symmetry~\cite{dash2019} and plot in
panels (b) and (c) of Fig.~\ref{fig:method-compare} the resulting differences
$\Delta E^{\rm PT2}$ between the PT2 energy corrections and $\Delta\sigma^2_{\rm
CI}$ between the CI variances of the ground and excited states (black symbols
labeled \quotes{cmax}).  We find that the convergence of the ground state is faster
than that of the excited state: $\Delta E^{\rm PT2}$ decreases and $\Delta\sigma^2_{\rm
CI}$ increases over the whole range of expansion sizes considered (see also Table S3). The faster
convergence of the ground state is further reflected in the excitation energy
$\Delta E_{\rm CI}$, shows in panel (d), which also grows with the expansions'
size. This is consistent with the observation that partitioning the determinants
in each expansion based on the relative contribution to each state gives an
insufficient number of ``excited-state'' determinants, namely, about 1.3-1.6 times
more determinants contributing to the excited than to the ground state (see Table
S4) compared to a significantly higher ratio of about 2.4-2.9 in our
previous calculations with matched PT2 energy corrections for formaldehyde and
thioformaldehyde~\cite{dash2019}.  Therefore, all indications are that the description
of the states is increasingly unbalanced in favor of the ground state as
the number of determinants gets larger, at least within the range explored.

To achieve  a more even description of the two states, we therefore modify the
selection criterion by introducing the state-average weights, $w_n^{\rm SA}$
(Eq.~\ref{eq:wSA}), and place a larger weight on the excited state. We show the
resulting difference of the ground- and excited-state PT2 energy corrections and
CI variances for sets of expansions generated with different choices of the
state-average weights (\quotes{cmax $\times$ SA}) in Fig.~\ref{fig:method-compare}.
The modified scheme clearly represents an improvement on the bare \quotes{cmax}
criterion and the use of 0.425 and 0.575 as $w_n^{\rm SA}$ weights for the ground
and the excited state, respectively, leads to nearly optimal matching of PT2
corrections and CI variances for all expansion sizes.

Partitioning the determinants based on their dominant relative contributions to
the two states reveals that the ratio between ``excited-state'' and ``ground-state''
determinants is now increased to about 1.9-2.1 as shown in Table S4. Furthermore,
an inspection of the CI energies $E_{\rm CI}$ obtained with both schemes, plotted
in panel (a) of Fig.~\ref{fig:vm-maugd}, shows that, for comparable sizes of the
expansion, the modified selection criterion slows down the convergence of the
ground state and, to a much lesser extent, speeds up the excited state.  The
CI excitation energy, compared in panel (d) with the CC3 value obtained in the
same basis, is also significantly reduced from about 6.6~eV with the \quotes{cmax}
criterion to 5.9~eV for the larger expansions considered here. For comparison, the
CC3/aug-cc-pVTZ excitation energy is 5.65~eV.  
Importantly, we find that the ideal
weights depend rather weakly on the orbital basis employed in the determinant
selection (see Fig.~S1) and also on the basis set (see Figs.~S2 and S3), being
always very close to 0.4 and 0.6.
Finally, we note that, while we set here the state-average weights manually, one can in principle devise
a scheme where the values of $w_n^{\rm SA}$
are dynamically adjusted during the CIPSI iterations to enforce the iso-PT2 condition as
closely as possible.

\subsection{Impact of selection on QMC-CIPSI excitation energies}
\label{sec:chap5-vexc}

\begin{figure}[htb!]
\includegraphics[width=1.0\columnwidth]{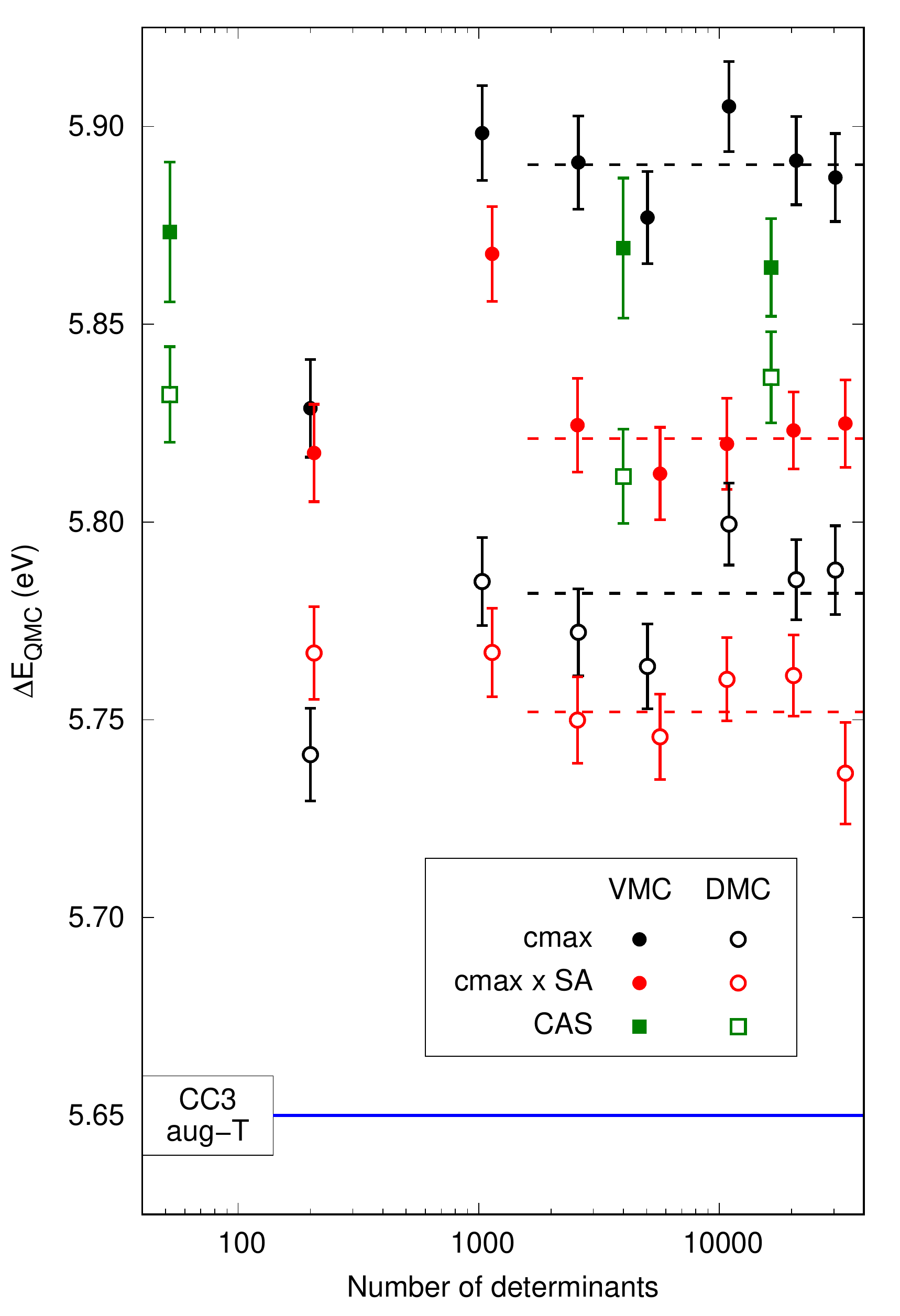}
\caption{
VMC (filled) and DMC (empty symbols) vertical excitation energies $\Delta E_{\rm
QMC}$ of thiophene versus the number of determinants for CIPSI determinantal
expansions generated with the \quotes{cmax} (black) and \quotes{cmax $\times$ SA}
(red) schemes with state-average weights 0.425 and 0.575.  The horizontal dashed
lines are drawn as guides and correspond to the average of the last 5 values of
the VMC or DMC excitation energies.  We also show VMC and DMC results for the
CAS determinantal expansions (6,5), (10,9), and (10,14) in order if increasing
number of determinants.
The maug-cc-pVDZ basis set is used for all QMC calculations.}
\label{fig:vm-maugd}
\end{figure}

In Fig.~\ref{fig:vm-maugd}, we illustrate the impact of the CIPSI selection on the
VMC and DMC vertical excitation energies of thiophene computed with the corresponding
Jastrow-CIPSI wave functions fully optimized at the VMC level. To this aim, we
consider the CIPSI expansions generated with the old \quotes{cmax} and the new
\quotes{cmax $\times$ SA} scheme and the ideal weights.

For all but the smallest expansion sizes, the VMC excitation energies 
corresponding to the \quotes{cmax} determinantal components settle around
about 5.9~eV.  Performing DMC calculations for these Jastrow-CIPSI wave functions
decreases the excitation energy, which remains however more than 0.1~eV higher
than the CC3 reference. Therefore, while both VMC and DMC substantially improve
on the starting CIPSI excitation energy, which is as high as 6.6~eV for the lagest
expansions considered here (see Fig.~\ref{fig:method-compare}), the bias
of the CIPSI selection towards the ground state is reflected in the QMC overestimation
of the excitation energy.

When we use the expansions obtained with the \quotes{cmax $\times$ SA} reweighting
scheme, we observe that the convergence of the ground state is somewhat slowed
down with respect to \quotes{cmax} expansions of comparable size, while the VMC
excited-state energies are largely unaffected (see Table S1). This leads to reduced
VMC excitation energies, which quickly converge to about 5.8~eV.  The VMC correction
on the CIPSI excitation energy is now smaller and the same holds for the improvement
of DMC upon VMC.

For comparison, in Fig.~\ref{fig:vm-maugd}, we also plot the QMC excitation energies
computed with Jastrow-Slater wave functions built with three different CAS expansions:
the minimal CAS(6,5), a CAS(10,9) correlating two additional occupied $\sigma$ and
two unoccupied $\sigma^{*}$ orbitals on the C$-$S bonds, and a CAS(10,14) that
further includes the five 3$d$ orbitals of the S atom.  For the largest active
space, the expansions are truncated with a threshold of 5$\times$10$^{-4}$ on the
CSF coefficients and the union of the CSFs of the ground and excited states is
retained.  While increasing the active space lowers the QMC total energies (see
Table S1), in all three cases, the VMC excitation energies are compatible within
statistical errors and rather comparable to their DMC counterparts, which remain
higher the DMC values obtained with the CIPSI expansions and either criterion.

\subsection{Best-quality QMC excitation energy}

\begin{figure}[h!]
\includegraphics[width=1.0\columnwidth]{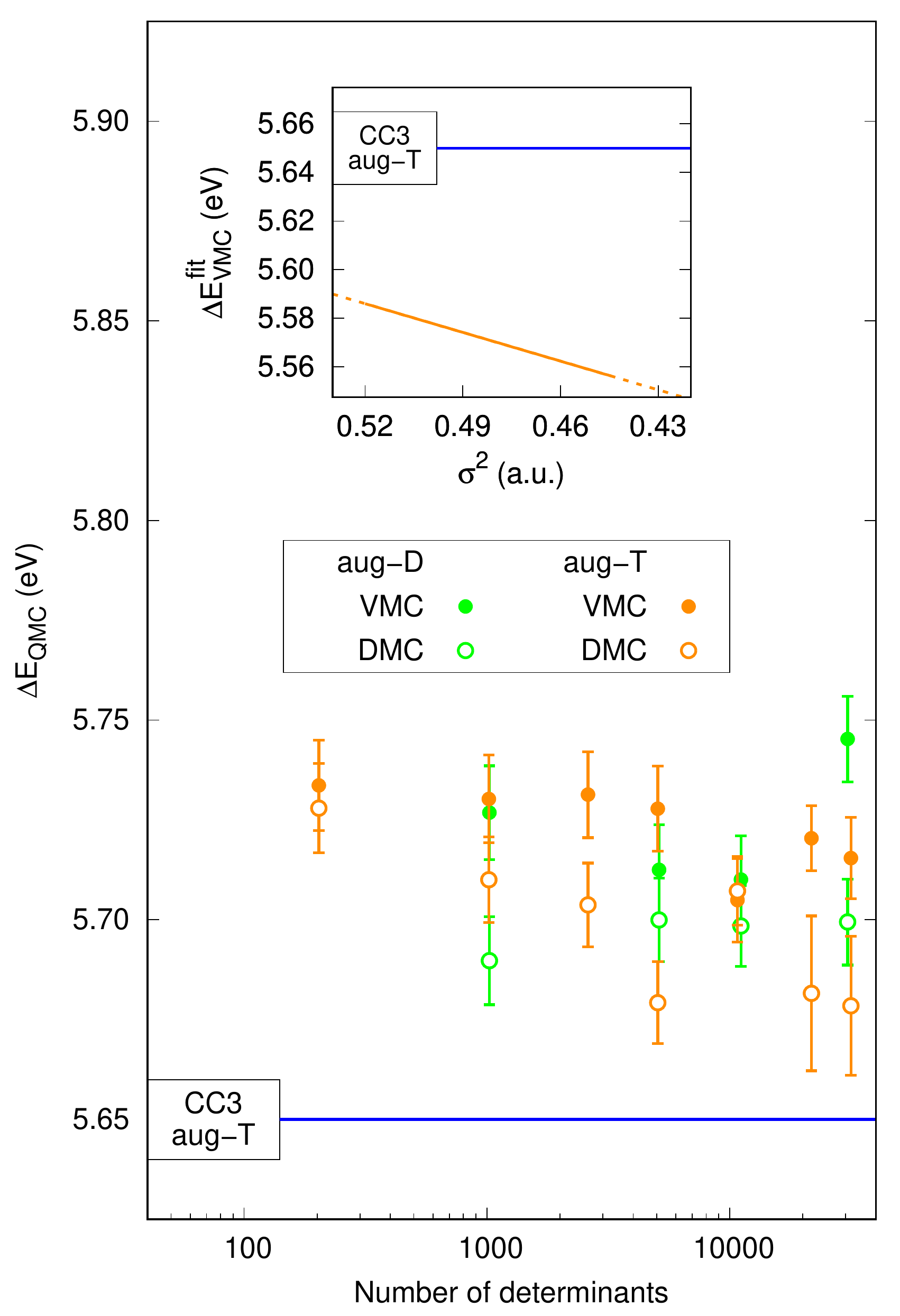}
\caption{
VMC (filled) and DMC (empty circles) vertical excitation energies $\Delta E_{\rm QMC}$ 
of thiophene versus the number of determinants for CIPSI expansions
generated with the aug-cc-pVDZ (aug-D) and aug-cc-pVTZ (aug-T) basis sets and the
\quotes{cmax $\times$ SA} scheme.  The VMC excitation energy with the aug-T basis
is also estimated as the difference 
$\Delta E_{\rm VMC}^{\rm fit}=E_{\rm ES}^{\rm fit}(\sigma^2)-E_{\rm GS}^{\rm fit}(\sigma^2)$ 
of the fits of the energies against the VMC variance of the two states (inset; the line 
of the fit difference is solid over the range of variances covered by the 
wave function used, with lower values of $\sigma^2$ corresponding to larger numbers 
of determinants).
}
\label{fig:vm-ta}
\end{figure}

\begin{table*}[htb!]
\centering
\caption{VMC and DMC ground- and excited-state energies (a.u.), and vertical excitation energies (eV) of thiophene for
increasing CIPSI expansions in fully optimized Jastrow-Slater wave functions and different basis sets.
The \quotes{cmax $\times$ SA} selection criterion is used. 
}
\label{tab:vexc}
\begin{tabular}{rrrcccccccccccccccc}
\hline
No.\ det &  No.\ param &&   \multicolumn{3}{c}{VMC}                          &&   \multicolumn{3}{c}{DMC}   \\
\cline{4-6}\cline{8-10}
         &             &&   E(GS)        &    E(ES)        & $\Delta$E       &&   E(GS)       & E(ES)         & $\Delta$E  \\
\hline
\multicolumn{10}{c}{aug-cc-pVDZ} \\[0.5ex]
1023     & 3084        &&  -35.28511(30) & -35.07512(31)   & 5.714(12)       && -35.35234(29) & -35.14325(29) & 5.690(11) \\
5116     & 6389        &&  -35.29685(29) & -35.08692(30)   & 5.712(11)       && -35.35678(27) & -35.14731(27) & 5.700(10) \\
11122    & 9412        &&  -35.30335(28) & -35.09350(29)   & 5.710(11)       && -35.35884(26) & -35.14943(26) & 5.698(10) \\
30615    & 16370       &&  -35.31022(28) & -35.09988(28)   & 5.724(11)       && -35.36062(28) & -35.15117(28) & 5.699(11) \\[1ex]
\multicolumn{10}{c}{aug-cc-pVTZ} \\[0.5ex]
1019     &  7442       &&   -35.29270(28) &  -35.08212(29)  & 5.730(11)      && -35.35744(28) & -35.14760(28) & 5.710(11) \\
5051     &  14656      &&   -35.30375(27) &  -35.09326(28)  & 5.728(11)      && -35.36164(26) & -35.15294(27) & 5.679(10) \\
10755    &  19674      &&   -35.30913(27) &  -35.09948(27)  & 5.705(10)      && -35.36438(22) & -35.15464(22) & 5.707(09) \\
31581    &  33363      &&   -35.31600(26) &  -35.10596(27)  & 5.715(10)      && -35.36534(46) & -35.15666(44) & 5.678(17) \\[1ex]
\hline
\multicolumn{4}{l}{CC3/aug-cc-pVDZ (BFD)}  & &  &&  &  & 5.678 \\
\multicolumn{4}{l}{CC3/aug-cc-pVTZ (BFD)}  & &  &&  &  & 5.65 \\
\hline
\end{tabular}
\end{table*}

The use of the \quotes{cmax$\times$SA} criterion to generate iso-PT2 CIPSI expansions
has the desired effect of yielding a more balanced description of the two states
also at the QMC level. Furthermore, the calibration of the state-average weights
we have carried out for the small maug-cc-pVDZ basis helps us in selecting an
appropriate range of values also when constructing wave functions with other basis
sets.

In particular, since the use of a maug-cc-pVDZ basis set is not sufficient for
an accurate treatment of the vertical excitation energy of thiophene, we generate new CIPSI
expansions with both the aug-cc-pVDZ and aug-cc-pVTZ basis sets, using the
\quotes{cmax $\times$ SA} scheme, and find that the choice of weights, 0.4 and
0.6, leads to well matched PT2 contributions and CI variances for both basis sets.
As shown in Fig.~\ref{fig:vm-ta}, the resulting VMC and DMC excitation energies
are red-shifted relative to comparable expansions in the smaller maug-cc-pVDZ
basis, and lie within less than 0.05~eV of our best CC3 reference value for both
basis sets.

Finally, we also estimate the vertical excitation energies by matching the VMC
variances for the wave functions~\cite{robinson2017,pineda2018}. To this aim, we
linearly fit the VMC ground- and excited-state energies separately against their
corresponding VMC variances $\sigma^2$ and then compute the excitation energy as
$E_{\rm ES}^{\rm fit}(\sigma^2) - E_{\rm GS}^{\rm fit}(\sigma^2)$. As
shown in the inset of Fig.~\ref{fig:vm-ta}, we find that the estimate of the
excitation energy falls below the CC3 reference and decreases with increasing
number of determinants, deviating for the largest expansions by about 0.1~eV.
Therefore, while the adoption of the iso-PT2 \quotes{cmax $\times$ SA} reweighting
scheme yields a more consistent estimate of the excitation energies, we observe a
somewhat less predictable behavior when matching the VMC variances.

\subsection{Optimal ground- and excited-state structures}
\label{sec:chap5-optgeo}

To optimize the structure of thiophene in the
ground and excited states in VMC, the maug-cc-pVDZ basis set 
is found to be sufficiently accurate as verified with the use of the 
aug-cc-pVDZ basis set in Table S7, and we will therefore proceed with 
this cheaper basis set to determine the structural parameters of the minima.

For the ground state, we start from the geometry and wave functions previously employed
to compute the vertical excitation energies of Table~\ref{tab:vexc}, for selected
values of the number of determinants.  For all chosen expansion sizes, we obtain
very accurate geometries with differences of about 5 m{\AA} in the bond lengths
with respect to the reference values as shown in Table~\ref{tab:thio-geo}.

\begin{table*}[htb!]
\centering
\caption{Optimal VMC ground- and excited-state bond lengths (\AA) and angles (deg) of thiophene
using \quotes{cmax $\times$ SA} CIPSI expansions and the maug-cc-pVDZ basis. 
}
\label{tab:thio-geo}
\begin{tabular}{lcrrccccccll}
\hline
State & $\delta$E$_{\rm PT2}$ & No. det  & No. param     &     C$-$C       &   C$=$C       &   C$-$S       &   $\delta_{\rm CCCS}$  \\
\hline
GS & -0.67 & 1037  & 2002 & 1.4281(2) & 1.3662(1) & 1.7201(2) & 0.06(5) \\ 
    & -0.64 & 2614  & 2823 & 1.4282(2) & 1.3681(2) & 1.7202(1) & 0.09(3) \\ 
    & -0.59 & 5605  & 4106 & 1.4290(2) & 1.3669(1) & 1.7218(1) & -0.08(3) \\ 
    & -0.54 & 11003 & 6326 & 1.4279(4) & 1.3676(4) & 1.7223(4) & 0.05(4) \\
\multicolumn{4}{l}{CASPT2$^a$}       & 1.430 & 1.372 & 1.720 & 0.00 \\
\multicolumn{4}{l}{CCSD(T) (BFD)$^b$}& 1.425 & 1.368 & 1.717 & 0.00 \\
\multicolumn{4}{l}{CCSD(T) (FC)$^b$} & 1.430 & 1.372 & 1.728 & 0.00 \\
\multicolumn{4}{l}{CC3 (FC)$^b$}     & 1.430 & 1.372 & 1.729 & 0.00 \\
%
ES & -0.67  &   1663     & 3630 &    1.4396(7)    &  1.4161(6)    &  1.7626(5)    &  24.84(2)  \\
    & -0.63  &   3752     & 4957 &    1.4388(7)    &  1.4151(3)    &  1.7655(4)    &  25.58(3)  \\
    & -0.59  &   8304     & 7271 &    1.4383(1)    &  1.4144(2)    &  1.7709(1)    &  25.75(2)  \\
    & -0.546  &   15815    & 10278 &    1.4422(7)    &  1.4112(4)    &  1.7725(6)    &  26.01(7)  \\[1ex]
\multicolumn{4}{l}{CASPT2$^a$}    & 1.448       & 1.423/1.416 & 1.782/1.778 & 26.7 \\
\multicolumn{4}{l}{ADC(2)$^c$}    & 1.422       & 1.419       & 1.796       & 28.2 \\
\multicolumn{4}{l}{DFT/MRCI$^d$}  & 1.436       & 1.394       & 1.799       & 27.4 \\
\hline
\multicolumn{8}{l}{$^a$MS-CASPT2(10,8)/6-31G** from Ref.~\cite{stenrup2012}.} \\
\multicolumn{8}{l}{$^b$CC/aug-cc-pVTZ either all-electron (FC) or pseudotpotential (BFD).}\\
\multicolumn{8}{l}{$^c$ADC(2)/cc-pVTZ from Ref.~\cite{prlj2015b}.}\\
\multicolumn{8}{l}{$^d$DFT/MRCI/TZVP from Ref.~\cite{salzmann2008}.}
\end{tabular}
\end{table*}

\begin{figure}[htb!]
\centering
\includegraphics[width=0.85\columnwidth]{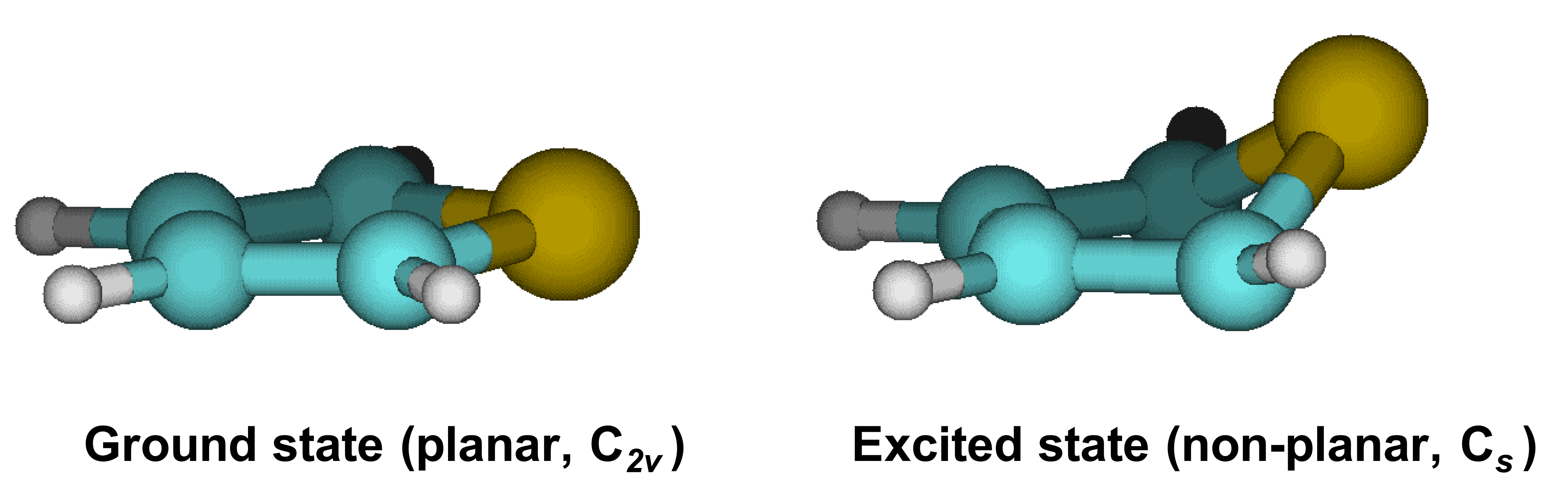}
\caption{Optimal ground- and excited-state structures of thiophene.}
\label{fig:exc-geometry}
\end{figure}

The excited-state global minimum is the S-puckered structure of symmetry C$_s$
shown in Fig.~\ref{fig:exc-geometry} as obtained in previous studies employing the
ADC(2)~\cite{prlj2015b}, DFT/MRCI~\cite{salzmann2008,grimme_1999}, and
MS-CASPT2~\cite{stenrup2012} approaches. It is interesting to note that TDDFT
predicts instead a C-puckered minimum with the S-puckered structure actually being
a transition state~\cite{prlj2015}.  Since the excited state has the same symmetry
as the ground state also when the molecular symmetry is lowered from C$_{2v}$ to
C$_s$, we need to follow the path of steepest descent for the second root in
optimizing the excited-state structure.  We do not impose any symmetry
constraints and start from a slightly distorted geometry, where we re-optimize
the same wave functions of Table~\ref{tab:vexc} selected for the ground-state
structural optimization.  In the subsequent steps, we generate iso-PT2 CIPSI wave
functions along the whole minimization path for both the ground and the excited
state.

The final converged structural parameters are shown in Table~\ref{tab:thio-geo}:
the optimized geometry is S-puckered in agreement with previous correlated
calculations, with an elongation of all bonds, in particular of the former C$=$C
and the C$-$S bonds.  The bond lengths and angles obtained with Jastrow-CIPSI  wave
functions are within 10 m{\AA} and 1$^{\circ}$ of the corresponding CASPT2 values,
respectively.

\section{Conclusions}
\label{sec:chap5-summary}

In this work, we present a systematic investigation on how to obtain balanced CIPSI
expansions for multiple states of the same symmetry and corresponding high-quality
QMC excitation energies and optimal excited-state geometries.  We focus on a case,
the lowest-energy bright state of thiophene, which is characterized by a significant
multi-reference character.

To this aim, we adapt here the CIPSI selection scheme to treat multiple states of the same 
symmetry with wave functions expressed on a common set of determinants, by introducing additional 
weights in the energy threshold for the selection step.  
This enables us to obtain expansions for the two states with the same PT2 energy corrections
and CI variances that is, similar estimated errors with respect to the full 
configuration interaction limit.
Importantly, we find that 
the modification introduced
in the selection scheme are largely independent of the basis set size
and orbital choice as the rate of convergence of the energy appears to be 
governed by the intrinsic multi-reference character of the excited state.

In practice, for thiophene, the new criterion slows down the convergence
of the ground state at every CIPSI iteration and, to a lesser extent, accelerates the one
of the excited state, making the quality of the two CI wave functions
more similar.  
Using these expansions as determinantal components in Jastrow-Slater wave functions
leads to DMC excitation energies 
within 0.05(2)~eV of the theoretical best estimate available, already when
the expansions comprise only about 5000 determinants.
With these Jastrow-CIPSI wave functions, structural relaxation in the ground state
yields VMC bond lengths which are compatible with the reference values to better than 0.01 
\AA. Following the second root in the geometry optimization while maintaining an iso-PT2 
description of the ground and excited states, 
we obtain optimal VMC excited-state bond lengths and angles, which are within less than 0.01 \AA\
and 1$^{\circ}$ of the best available estimates, respectively. 

In summary, also in a case like thiophene
where a reliable treatment fully at the CIPSI level is quite 
demanding, we are able to generate balanced Jastrow-Slater wave functions 
for multiple states and determine accurate excited-state properties in QMC, 
using relatively short CIPSI expansions.  Therefore,
when increasing the number of electrons or the size of the basis set,
we expect that QMC in combination with compact and balanced CIPSI expansions 
will remain a viable route to deliver 
high-quality excited-state potential energy surfaces in the domain of
applications where selected configuration interaction becomes
instead intractable.\\

\section*{Acknowledgment}

This work is part of the Industrial Partnership Programme (IPP) ``Computational
sciences for energy research'' of the Netherlands Organisation for Scientific
Research (NWO-I, formerly FOM). This research programme is co-financed by Shell
Global Solutions International B.V. This work was carried out on the Dutch
national supercomputer Cartesius with the support of SURF Cooperative.
The authors declare no competing financial interest.

\section*{Suppinfo}

CIPSI energies, PT2 energy corrections, and CI variances; 
VMC and DMC ground- and excited-state energies and vertical excitation energies 
computed with the maug-cc-pVDZ and aug-cc-pVTZ basis sets; dependence of the DMC energies on the time step; dependence of
the $w_{n}^{\rm SA}$ weights on the choice of orbitals and basis set; partitioning of the CSFs between the two states;
comparison with other levels of theory; ground- and excited-state VMC geometries obtained with the aug-cc-pVDZ basis set; 
ground-state CC3 geometry.  

\bibliography{thiophene-cipsi}

\begin{thebibliography}{50}
\bibitem [{BFD()}]{BFD_H2013}%
  \BibitemOpen
  \href@noop {} {}\bibinfo {note} {For the hydrogen atom, we use a more
  accurate BFD pseudopotential and basis set. Dolg, M.; Filippi, C., private
  communication}\BibitemShut {NoStop}%
\bibitem [{Jas()}]{Jastrow}%
  \BibitemOpen
  \href@noop {} {}\bibinfo {note} {As Jastrow factor, we use the exponential of
  the sum of two fifth-order polynomials of the electron-nuclear and the
  electron-electron distances, respectively, and rescale the inter-particle
  distances as $R=(1-\exp(-\kappa r))/\kappa$ with $\kappa$ set to 0.6 a.u. We
  employ different electron-nucleus Jastrow factors to describe the correlation
  of an electron with C, S, and H. The total number of free parameters to be
  optimized in the 2-body Jastrow factor is 17 for thiophene.}\BibitemShut
  {Stop}%
\end{thebibliography}
\end{document}